\documentclass[12pt,a4paper]{article}
\usepackage{epsfig}
\begin{document}
\title{Associated production of the scalars and new gauge bosons from a little
Higgs model at the $LHC$}
\author{Chong-Xing Yue, Nan Zhang, Li Ding, and Shi-Hai Zhu\\
{\small Department of Physics, Liaoning  Normal University, Dalian
116029, P. R. China}
\thanks{E-mail:cxyue@lnnu.edu.cn}}
\date{\today}

\maketitle

\begin{abstract}
The littlest Higgs model with T-parity ($LHT$ model) predicts the
existence of the T-odd scalars ($\phi^{\pm}$, $\phi^{0}$, and
$\phi^{p}$). We consider production of these new particles
associated with T-odd gauge bosons at the $LHC$. We find that the
partonic process $q\overline{q'}\rightarrow\phi^{+}B_{H}$ can
generate a number of the characteristic signal events with a charged
lepton and large missing energy at the $LHC$.
 \vspace{1.8cm}

PACS number:12.60.Cn, 14.80.Cp, 14.70.Pw

\end{abstract}

\newpage
\noindent{\bf 1. Introduction}

The mechanism of electroweak symmetry breaking $(EWSB)$ and origin
of the fermion mass remain unknown in the current particle physics
despite the success of the standard model $(SM)$ tested by high
energy experimental data. In next year, the $CERN$ Large Hadron
Collider $(LHC)$ will begin operation, which has an increase of a
factor of seven in energy and a factor of 100 in luminosity over the
Fermilab Tevatron. The $LHC$ can be seen as the machine extremely
well suit to the study of these problems and is expected to directly
probe possible new physics beyond the $SM$ up to a few TeV, which
might provide some striking evidence of new physics.

Little Higgs theory [1] is proposed as an interesting solution to
the so called hierarchy problem of the $SM$ and can be regarded as
one of the important candidates for new physics beyond the $SM$.
Among these models, the littlest Higgs $(LH)$ model [2] is one of
the simplest and phenomenological viable models, which has all
essential features. Most phenomenological analysis about the
original little Higgs models are given in the context of the $LH$
model [3, 4]. They have shown that some of new particles predicted
by the $LH$ model can generate characteristic signatures at present
and in future collider experiments.

It is well known that the $LH$ model suffers from severe constraints
from precision electroweak measurement, which could only be
satisfied by fine-tuning the model parameters [5]. To avoid this
serious problem, a new discrete symmetry called T-parity has been
introduced. The $LH$ model with T-parity $(LHT$ model) [6] is one of
the attractive little Higgs models, which satisfies the electroweak
precision data in most of the parameter space. In the $LHT$ model,
all of the $SM$ particles are assigned with an T-even parity, while
all of the new particles are assigned with an T-odd parity, except
for the little Higgs partner of the top quark. As a consequence, all
of the T-odd particles can only be generated in pairs. The
electroweak precision data allow for a relatively low value of new
particle mass scale $f$ $\sim$ 500$GeV$ [7, 8]. Thus, the $LHT$
model can produce rich phenomenology at present and in future high
energy experiments [9, 10, 11].

Most of the previous works on studying the phenomenology of the
$LHT$ model focus their attention on the new fermions (T-odd
fermions or T-even heavy top quark $T^{+}$) and new gauge bosons
$(W_{H}^{\pm}, B_{H}, Z_{H})$. In this letter, we will discuss
single production of the scalars ($\phi^{0}$, $\phi^{p}$, and
$\phi^{\pm}$) predicted by the $LHT$ model via possible processes at
the $LHC$ with the center-of-mass (c. m.) energy $ \sqrt{S}=14 TeV$.
In section 2, we briefly summarize the mass and coupling formula
related to our calculation. Single production of the new scalars in
association with new gauge bosons at the $LHC$ are discussed in
section 3. The relevant phenomenological analysis are also given in
this section. In section 4, we give our conclusions.

 \noindent{\bf 2. The relevant mass and coupling formula}

Similar to the $LH$ model, the $LHT$ model [6] is based  on an
$SU(5)/SO(5)$ global symmetry breaking pattern. A subgroup
$[SU(2)_{1}\times U(1)_{1}]\times[SU(2)_{2}\times U(1)_{2}]$ of the
$SU(5)$ global symmetry is gauged, and at the scale $f$ it is broken
into the $SM$ electroweak symmetry $SU(2)_{L}\times U(1)_{Y}$.
T-parity is an automorphism that exchanges the $[SU(2)_{1}\times
U(1)_{1}]$ and $[SU(2)_{2}\times U(1)_{2}]$ gauge symmetries.
Consequently, four of the 14 Nombu-Goldstone$(NG)$ bosons are eaten
by the T-odd heavy gauge bosons $(W_{H}^{\pm}, Z_{H}, B_{H})$
associated with the broken symmetry. The remaining $NG$ bosons
decompose into a T-even doublet $H$, considered to be the $SM$ Higgs
doublet, and a complex T-odd $SU(2)$ triplet $\Phi$. There is a
relation between the triplet and doublet Higgs boson masses, which
is approximately expressed as:
\begin{equation}
M_{\Phi}=\frac{\sqrt{2}m_{H}}{\nu}f,
\end{equation}
where $\nu$=246GeV is the electroweak scale and $f$ is the scale of
the gauge symmetry breaking of the $LHT$ model.

After $EWSB$, at the order of $\nu^{2}/f^{2}$, the mass expressions
of the T-odd heavy gauge bosons $B_{H}$, $Z_{H}$ and $W_{H}$ are
given by
\begin{equation}
M_{B_{H}}=\frac{g'f}{\sqrt{5}} [1-\frac{5\nu^{2}}{8f^{2}}],
\hspace{0.5cm}M_{Z_{H}} \approx
M_{W_{H}}=gf[1-\frac{\nu^{2}}{8f^{2}}].
\end{equation}
Here $g'$ and $g$ are the $SM$ $U(1)_{Y}$ and $SU(2)_{L}$ gauge
coupling constants, respectively. Because of the smallness of $g'$,
the T-odd gauge boson $B_{H}$ is the lightest T-odd particle, which
can be seen as an attractive dark matter candidate [9].

To implement T-parity in the fermion sector, one introduces two
$SU(2)$ fermion doublets for each $SM$ fermion doublet [6, 7, 8].
The T-even linear combination is associated with the $SM$ $SU(2)$
doublet, while the T-odd combination is associated with its T-parity
partner. Its mass is at order of the symmetry breaking scale $f$.
Assuming universal and flavor diagonal Yukawa coupling constant $k$,
the masses of the T-odd fermions $u_{-}$ and $d_{-}$ can be
approximately written as:
\begin{equation}
M_{u_{-}}\approx \sqrt{2}kf(1-\frac{\nu^{2}}{8f^{2}}),
\hspace{0.5cm}M_{d_{-}}\approx \sqrt{2}kf.
\end{equation}

The couplings of these new particles to the $SM$ particles (T-even),
which are related our calculation, can be approximately written as
[8, 11]:
\begin{eqnarray}
&\phi^{0}Z^{\mu}B_{H}^{\nu}&:\frac{ig'^{2}}{2\sqrt{2}S_{W}}\frac{\nu^{2}}{f}g^{\mu\nu},
\hspace{0.5cm}
\phi^{0}Z^{\mu}Z_{H}^{\nu}:-\frac{ig^{2}}{2\sqrt{2}C_{W}}\frac{\nu^{2}}{f}g^{\mu\nu};
\\
&\phi^{p}W_{H}^{-\mu}W^{+\nu}&:\frac{g^{2}}{3\sqrt{2}}\frac{\nu^{2}}{f},g^{\mu\nu},
\hspace{1.1cm}
\phi^{p}u^{i}\overline{u}_{-}^{j}:\frac{k}{12}\frac{\nu^{2}}{f^{2}}(V_{Hu})_{ij}P_{L};
\\
&\phi^{+}B_{H}^{\mu}W^{-\nu}&:\frac{igg'}{4}\frac{\nu^{2}}{f}g^{\mu\nu},\hspace{0.58cm}
\phi^{+}W_{H}^{-\mu}Z^{\nu}:-\frac{ig^{2}}{12C_{W}}(1+2S_{W}^{2})\frac{\nu^{2}}{f}g^{\mu\nu};
\\
&\phi^{+}W_{H}^{-\mu}\gamma^{\nu}&:\frac{ieg}{6}\frac{\nu^{2}}{f}g^{\mu\nu},\hspace{0.66cm}
\phi^{+}Z_{H}^{\mu}W^{-\nu}:\frac{i5g^{2}}{12}\frac{\nu^{2}}{f}g^{\mu\nu};
\\
&Z_{H}^{\mu}d_{-}^{i}\overline{d^{i}}&:\frac{ig}{2}(V_{Hd})_{ij}\gamma^{\mu}P_{L},
\hspace{0.28cm}
B_{H}^{\mu}d_{-}^{i}\overline{d^{i}}:\frac{ig'}{10}(V_{Hd})_{ij}\gamma^{\mu}P_{L};
\\
&W_{H}^{+\mu}\overline{u_{-}^{i}}d^{j}&:\frac{ig}{\sqrt{2}}(V_{Hd})_{ij}\gamma^{\mu}P_{L},
\hspace{0.08cm}
\Phi^{+}d_{-}^{i}\overline{u}^{j}:\frac{k}{12}\frac{\nu^{2}}{f^{2}}(V_{Hu})_{ij}P_{L}.
\end{eqnarray}
Here $\phi^{+}$, $\phi^{-}$, $\phi^{0}$, and $\phi^{p}$ are the
components of the triplet scalar $\Phi$, which have the same mass at
the leading order. $k$ is the Yukawa-type coupling constant for the
mirror fermions. $S_{W}=\sin\theta_{W}$, $\theta_{W}$ is the
Weinberg angle. $(V_{Hu})_{ij}$ and $(V_{Hd})_{ij}$ are the matrix
elements of the CKM-like unitary mixing matrices $V_{Hu}$ and
$V_{Hd}$, respectively, which satisfy $V_{Hu}^{+}V_{Hd}=V_{CKM}$.

In the following section, we will use above Feynman rules to
calculate production of the T-odd scalars ($\phi^{p}$, $\phi^{0}$,
$\phi^{+}$ and $\phi^{-}$) in association with  new gauge bosons at
the $LHC$.

 \noindent{\bf 3. Associated production of the T-odd scalars and new
 gauge bosons at the \hspace*{0.6cm}\emph{\textbf{LHC}}}

 In the $LHT$ model [6, 8], T-parity explicitly forbids the tree level
 contributions from the new gauge bosons to the observables involving only
 the $SM$ particles and forbids the interactions that induce triplet vacuum expectation value
 ($VEV$). Corrections to electroweak observables are
 only generated at one loop level or beyond, which are naturally
 small. As a result, the electroweak precision data allow for
 a relatively low value of the symmetry breaking scale
 $f\sim500GeV$ [7]. Thus, the new particle mass might be relatively
 low, which induces that the T-partners of the $SM$ particles can be
 copious produced at the $LHC$ as well as future $e^{+}e^{-}$ linear
 collider $(ILC)$ [8, 10, 11]. In this section, we will consider
 production of the T-odd scalars in association with the T-odd
 new gauge bosons at the $LHC$.
\vspace{-2.0cm}
 \begin{figure}[htb]
\begin{center}
\epsfig{file=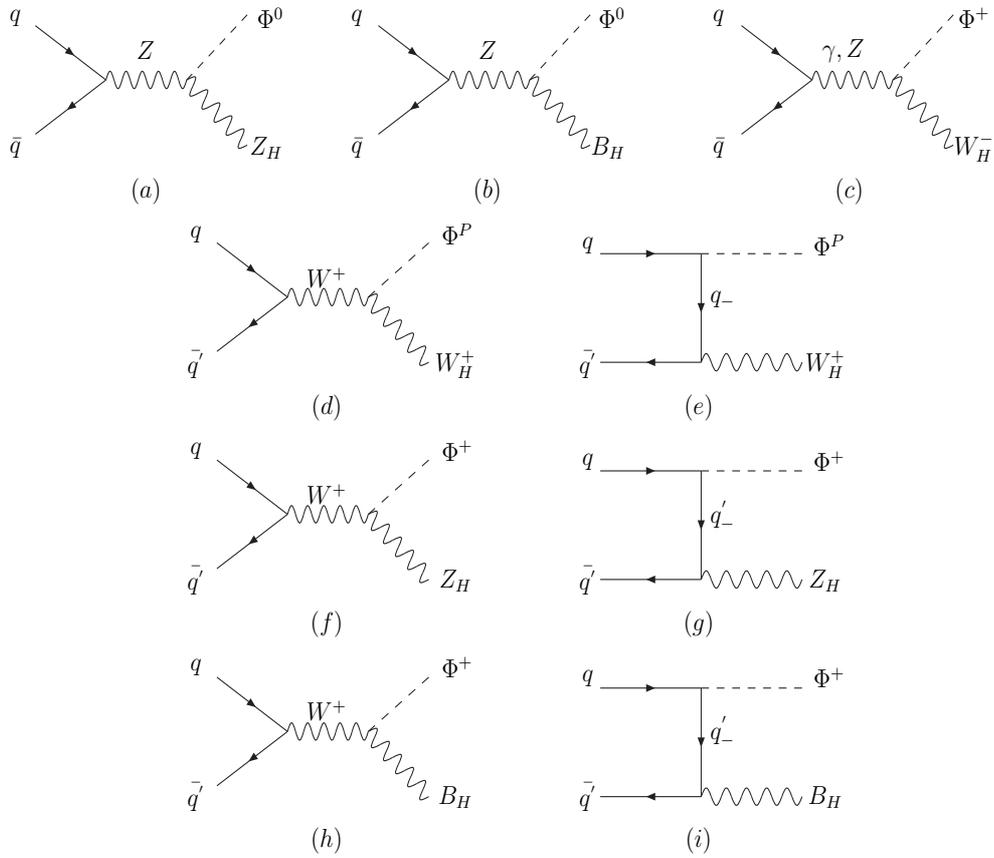,width=450pt,height=690pt} \vspace{-10.9cm}
 \caption{Feynman diagrams for the production of the new scalars at the $LHC$.}
 \label{ee}
\end{center}
\end{figure}

 From Feynman rules given in section 2, we can see that the T-odd
 scalars can be produced at the $LHC$ via the following partonic
 processes:
\begin{eqnarray}
q\overline{q}\rightarrow\phi^{0}Z_{H},\hspace{0.5cm}q\overline{q}\rightarrow
\phi^{0}B_{H},\hspace{0.5cm}q\overline{q}\rightarrow\phi^{-}W_{H}^{+};
\\
q\overline{q'}\rightarrow\phi^{p}W_{H}^{+},\hspace{0.5cm}q\overline{q'}
\rightarrow\phi^{+}Z_{H},\hspace{0.5cm}qq'\rightarrow\phi^{+}B_{H}.
\end{eqnarray}
The relevant Feynman diagrams are shown in Fig.1. One can see from
Fig.1 that the partonic processes $q\overline{q}\rightarrow
\phi{V_{H}}(V_{H}=Z_{H},B_{H}$, or $W_{H})$ can only be induced via
the s-channel exchange of the $SM$ gauge boson $Z$ or $\gamma$,
while the partonic processes $q \overline{q'} \rightarrow
\phi{V_{H}}$ can proceed by the s-channel exchange of the $SM$ gauge
bosons $W^{\pm}$ and the t-channel exchange of the T-odd mirror
fermion $q_{-}$.

Considering that the coupling $\phi qq_{-}$ is proportion to the
factor $\nu^{2}/f^{2}$, the t-channel contributions to the process
$q\overline{q'}\rightarrow\phi V_{H}$ should be suppressed by the
factor $\nu^{4}/f^{4}$. However, to completely consider production
cross section of the T-odd scalars at the $LHC$, we will include
both the s-channel and t-channel contributions in our numerical
calculation.

At the $LHC$, production of the Higgs boson in association with $W$
or $Z$ boson, $pp\rightarrow WH/ZH+X$, is one of the interesting
processes for detecting the $SM$ Higgs boson $H$ with a mass below
about 135$GeV$, where $H$ decays into $b\overline{b}$ final states
are dominate [12]. At leading order, production of a Higgs boson in
association with $W$ or $Z$ boson, $p p \rightarrow VH+X(V=W, Z)$,
proceeds through $q\overline{q'}$ annihilation
$q\overline{q'}\rightarrow V^{*}\rightarrow VH$ [13]. Production of
the T-odd scalar in association with a new gauge boson at the $LHC$
is similar with $WZ/ZH$ production. Thus, we can easily give the
production cross section $\sigma(\phi V_{H})$ for the process $p p
\rightarrow q\overline{q'}+X \rightarrow \phi V_H+X(V_{H}=W_{H},
Z_{H}, B_{H})$ at the $LHC$. In our numerical calculation, we will
use $CTEQ6L$ parton distribution functions ($PDF's$) [14] for the
quark distribution functions.

To obtain numerical results, we need to specify the relevant $SM$
input parameters, which are $\alpha_{e}=1/128.8, S_{W}^{2}=0.2315,
M_{W}=80.45GeV$, and $M_{Z}=91.187GeV$ [15]. Considering the
experimental constraints on the $SM$ Higgs boson mass $m_{H}$ [16],
we will take $m_{H}=120GeV$. Except these $SM$ input parameters, the
cross section $\sigma(\phi V_{H})$ for associated production of the
new scalar and the new gauge boson depends on the free parameters
$k$, $(V_{Hu})_{ij}$, $(V_{Hd})_{ij}$, and $f$. The matrix element
$(V_{Hd})_{ij}$ can be determined in flavor violating processes and
the matrix element $(V_{Hu})_{ij}$ is then determined through
$V_{Hu}=V_{Hd}V_{CKM}^{+}$. Following Ref.[11], we will take $k=1,
(V_{Hd})_{ij} (i=j)=\frac{1}{\sqrt{2}}$, and assume the scale
parameter $f$ as a free parameter in our numerical estimation.

Our numerical results are shown in Fig.2 and Fig.3. From these
figures, one can see that the production cross section quickly
decreases as $f$ increases. This is because the masses of the T-odd
scalars and gauge bosons increase as $f$ increases. Since the new
gauge boson $B_{H}$ is lighter than the new gauge boson $Z_{H}$ or
$W_{H}^{\pm}$, the production cross sections of the new scalars in
association with $B_{H}$ are generally larger than those for the new
scalars in association with gauge boson $Z_{H}$ or $W_{H}^{\pm}$.
However, in most of the parameter space of the $LHT$ model, all of
the production cross sections are small. Even for the process
$q\overline{q'}\rightarrow\phi^{+}B_{H}$, its production cross
section is only in the range of $1.014fb\sim0.024fb$ for $500GeV\leq
f\leq800GeV$. With the high luminosity option of the $LHC$, around
$300fb^{-1}$, there will be several and up to hundreds
$\phi^{+}B_{H}$ events.
 \vspace{-0.5cm}
\begin{figure}[htb]
\begin{center}
\epsfig{file=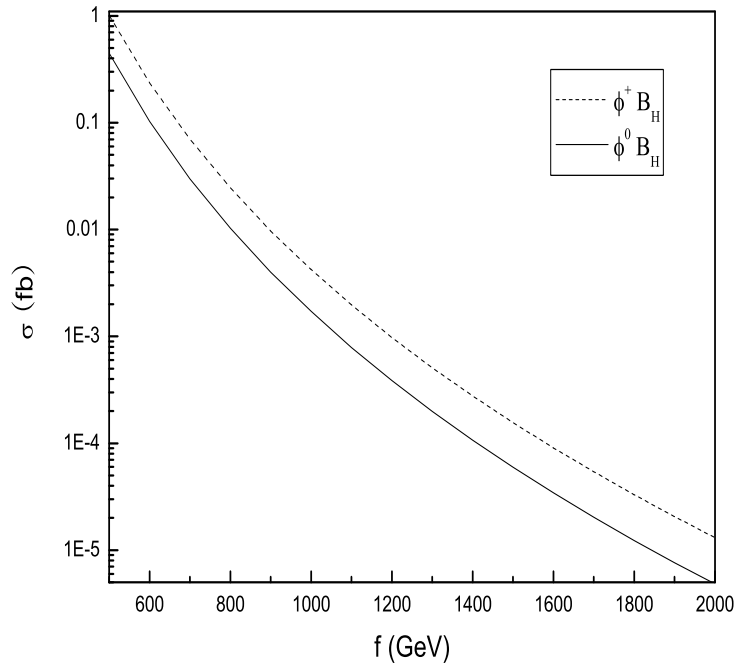,width=305pt,height=295pt} \vspace{-1.0cm}
 \caption{Production cross sections for the partonic processes $q\overline{q'}
 \rightarrow \phi^{+}B_{H}$ and $q\overline{q}\rightarrow \hspace*{1.8cm} \phi^{0}B_{H}$
 as functions of the scale parameter $f$.} \label{ee}
\end{center}
\end{figure}

From the discussion given in Sec.2, we can see that the charged
scalar $\phi^{+}$ mainly decays to $W^{+}B_{H}$ and there will be
$Br (\phi^{+}\rightarrow W^{+}B_{H})\approx1$ for $m_{H}=120GeV$ and
$k=1$[11]. In the case of $W^{+}$ decaying leptonically, the
partonic process $q\overline{q'}\rightarrow\phi^{+}B_{H}$ could give
rise to the distinct signal events with one charged lepton and large
missing energy, i.e.
$q\overline{q'}\rightarrow\phi^{+}B_{H}\rightarrow
W^{+}B_{H}B_{H}\rightarrow l^{+}\nu B_{H}B_{H}$.  Thus, as long as
$f\leq800GeV$, there will be several tens of observed $l^{+}\nu
B_{H}B_{H}$ events to be generated at the $LHC$ with $
\sqrt{S}=14TeV$ and a yearly integrated luminosity of
$\pounds=300fb^{-1} $. \vspace{-0.5cm}
\begin{figure}[htb]
\begin{center}
\epsfig{file=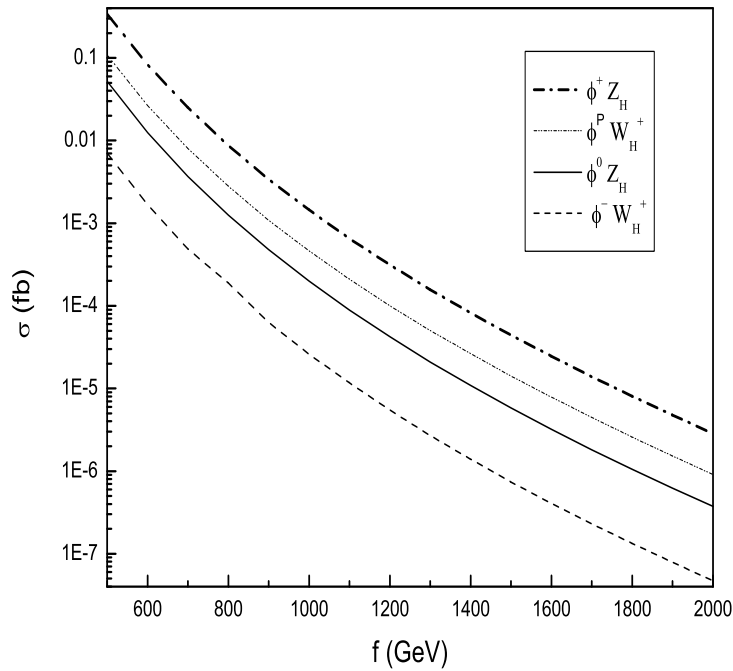,width=305pt,height=295pt} \vspace{-1.0cm}
 \caption{Same as Fig.2 but for the partonic processes $q\overline{q}\rightarrow
 \phi^{0}Z_{H}$, $q\overline{q}\rightarrow \phi^{-}W_{H}^{+}$,
 $q\overline{q'}\rightarrow \hspace*{1.7cm}\phi^{p}W_{H}^{+}$,
 and $q\overline{q'}\rightarrow \phi^{+}Z_{H}$.} \label{ee}
\end{center}
\end{figure}

\noindent{\bf 4. Conclusions }

The $LHT$ model is one of the attractive little Higgs models, which
provides a possible dark matter candidate. In this model, all
dangerous tree level contributions to electroweak observables are
forbidden by T-parity and hence the corrections to electroweak
observables are very small. So the low scale parameter $f
(f\geq500GeV)$ is allowed by the electroweak precision data and this
model might produce rich phenomenology at present and in future high
energy experiments.

In this letter, we consider production of the T-odd scalars
($\phi^{\pm}$, $\phi^{0}$, and $\phi^{p}$) in association with T-odd
gauge bosons at the $LHC$. Since all production processes are
proceed mainly via s-channel processes with highly virtual gauge
boson propagators, the production cross sections are small in most
of the parameter space of the $LHT$ model. However, the process
$q\overline{q'}\rightarrow \phi^{+}B_{H}$ can generate the
characteristic signal events with a charged lepton and large missing
energy. There will be several tens of observed events to be
generated at the $LHC$. Therefore, detecting the signature of
$\phi^{+}$ is challenging for the $LHC$,  depending on its
luminosity.

\vspace{0.5cm} \noindent{\bf Acknowledgments}

This work was supported in part by the Program for New Century
Excellent Talents in University(NCET-04-0290), and the National
Natural Science Foundation of China under the Grants No.10675057.

\end{document}